\documentclass[aps,prl,10pt,superscriptaddress,nofootinbib,twocolumn,preprintnumbers]{revtex4-1}
\usepackage[utf8]{inputenc}
\usepackage{amsmath,amssymb,amsfonts,graphicx,multirow,xspace}
\usepackage[dvipsnames]{xcolor}
\usepackage[colorlinks=true, linkcolor = blue]{hyperref} 
\hypersetup{colorlinks=true, linkcolor=BurntOrange, citecolor=BurntOrange, filecolor=BurntOrange, urlcolor=BurntOrange}

\DeclareUnicodeCharacter{2212}{-}

\newcommand{\delphill}{\ensuremath{\Delta \phi (\ell \ell)}\xspace}
\newcommand{\deletall}{\ensuremath{\Delta \eta (\ell \ell)}\xspace}
\newcommand{\mll}{\ensuremath{\text{m}_{\ell \ell}}\xspace}
\newcommand{\hnull}{\ensuremath{\text{H}_0}}
\newcommand{\halt}{\ensuremath{\text{H}_{\mathrm{toponium}}}}

\newcommand{\mg}{{\sc MG5\_aMC}\xspace}
\def\mgversion{\textsc{v3\_5\_7}\xspace}
\def\pythia{\textsc{Pythia}\xspace}
\def\pythiaversion{\textsc{v8.312}\xspace}
\def\rivet{\textsc{Rivet}\xspace}
\def\rivetversion{\textsc{v4.1.0}\xspace}
\def\rivetroutinecodeATLAS{\textsc{ATLAS\_2019\_I1759875}\xspace}
\def\rivetroutinecodeCMS{\textsc{CMS\_2018\_I1703993}\xspace}
\newcommand{\gitlink}{\href{https://github.com/keaveney/toponium-analysis/tree/main}{GitHub\xspace}}

\begin{document}

\title{Statistical Indications of Toponium Formation in Top Quark Pair Production}

\author{Benjamin Fuks}
\email{fuks@lpthe.jussieu.fr}
\affiliation{Laboratoire de Physique Théorique et Hautes Énergies (LPTHE), UMR 7589, Sorbonne Université et CNRS, 4 place Jussieu, 75252 Paris Cedex 05, France}

\author{Aminul Hossain}
\email{HSSAMI005@myuct.ac.za}
\affiliation{Department of Physics, University of Cape Town, University Avenue, Cape Town 7700, South Africa}

\author{James Keaveney}
\email{james.keaveney@uct.ac.za}
\affiliation{Department of Physics, University of Cape Town, University Avenue, Cape Town 7700, South Africa}
\affiliation{School of Physics, University College Dublin, Belfield, Dublin 4, Ireland}

\date{\today}
\begin{abstract}
  We present an analysis of six differential cross-section measurements of top-quark pair production in the dilepton channel from the ATLAS and CMS experiments. The data are compared to state-of-the-art QCD predictions with and without the inclusion of toponium formation effects. This contribution is modelled via a re-weighting of fixed-order matrix elements using the Green’s function of the non-relativistic QCD Hamiltonian, and we employ a statistical model to quantify the preference of the data for the toponium hypothesis. All observables yield Bayes factors larger than unity, with two exceeding 20, yielding strong evidence for the toponium hypothesis in top-quark pair production at the LHC.
\end{abstract}

\maketitle

%%%%%%%%%%%%%%%%%%%%%%%%%%%%%%%%%%%%%%%%%%%%%%%%%%%%%%%%%%%%%%%%%%%%%%%%%%%%%%%%%%%%%%%%%%
%%%%%%%%%%%%%%%%%%%%%%%%%%%%%%%%%%%%%%%%%%%%%%%%%%%%%%%%%%%%%%%%%%%%%%%%%%%%%%%%%%%%%%%%%%
%% main text

\textit{\textbf{Introduction}} -- The production of top-quark pairs ($t\bar{t}$) at the Large Hadron Collider (LHC) provides a unique laboratory to probe both Standard Model (SM) and possible new interactions involving the top quark. In particular, the kinematic threshold region, where the invariant mass of the $t\bar{t}$ pair satisfies $m_{t\bar{t}} \approx 2\,m_t$, is highly sensitive to the top-quark mass $m_t$, width and electroweak couplings including the top Yukawa, as well as to the effects of hypothetical beyond-the-Standard-Model states with masses near $2 m_t$. However, this sensitivity is limited by the theoretical modelling (such as the truncation of the perturbative QCD series, the use of the narrow-width approximation for the top quark decays or the interference with single-top processes) and by the experimental uncertainties.

In the SM, top-quark pairs produced near threshold in a colour-singlet configuration can experience Coulombic gluon exchange, leading to an enhancement of the production cross section relative to the conventional, non-resonant process. This phenomenon can be interpreted as toponium formation whose effects are calculable using the Green's function of the non-relativistic QCD Hamiltonian~\cite{Fadin:1987wz, Fadin:1990wx, Hagiwara:2008df, Sumino:2010bv}. In particular, toponium formation is expected to modify both the invariant mass distribution and the angular correlations of the $t\bar{t}$ system and its decay products~\cite{Fuks:2021xje, Fuks:2024yjj}. In this context, precision fixed-order computations have been achieved for specific observables~\cite{Ju:2020otc, Garzelli:2024uhe, Nason:2025hix}, while recent development of Monte Carlo simulations including toponium effects complementarily enables predictions for a wide range of particle-level observables at the LHC~\cite{Fuks:2021xje, Fuks:2024yjj, Fuks:2025wtq}.

Thanks to the large LHC datasets, the excellent performance of the ATLAS and CMS detectors and advances in experimental and theoretical techniques, the inclusive and differential $t\bar{t}$ cross sections have been measured with percent-level precision. Persistent discrepancies between predictions and data have been observed at both parton and particle level. These occur when using calculations matching next-to-leading-order (NLO) matrix elements to parton showers as well as QCD computations at the next-to-next-to-leading order (NNLO) further including NLO electroweak corrections and threshold resummation~\cite{ATLAS:2019hau, ATLAS:2019zrq, CMS:2018adi, CMS:2024ybg}. A subset of these deviations occurs in kinematic regions sensitive to toponium contributions, and dedicated detector-level analyses by ATLAS and CMS further reported evidence consistent with the toponium hypothesis~\cite{ATLAS:2023fsd, CMS:2024pts, CMS:2024ybg, CMS:2024zkc, CMS:2025kzt, CMS:2025dzq, ATLAS:2025kvb}. However, it is still essential to establish whether the data consistently support the hypothesis of toponium formation and if room is left for new physics contributions.

In this letter, we present an analysis of six published particle-level differential $t\bar{t}$ cross section measurements in the dilepton channel, chosen for their sensitivity to toponium formation, the availability of conventional predictions at NNLO precision and that of public \textsc{Rivet} implementations yielding an exact emulation of the relevant observable definitions. We consider particle-level observables to reduce the model dependence of the data as opposed to parton-level measurements and allow for a straightforward comparison to theoretical predictions. We then perform a statistical analysis to quantify the evidence for toponium formation and estimate the related signal cross section. This work thus represents the first exploration of toponium effects at the particle level combining data from both ATLAS and CMS.

\medskip

\textit{\textbf{Technical Framework}} -- Each of the six analysed differential cross sections is normalised to unity, thereby suppressing uncertainties that affect the normalisation of the distribution in both the data and the theoretical predictions. Furthermore, since toponium formation has a negligible effect on the inclusive $t\bar{t}$ production rate \cite{Hagiwara:2008df}, this normalisation procedure enhances the relative sensitivity to the shape distortions induced by toponium formation. The first observable examined is a two-dimensional distribution of the invariant mass of the dilepton system (\mll) and the azimuthal angle between the two leptons (\delphill) measured by ATLAS~\cite{ATLAS:2019hau}. The remaining five observables correspond to one-dimensional distributions in \mll and \delphill measured by ATLAS~\cite{ATLAS:2019hau} and CMS~\cite{CMS:2018adi}, and the absolute difference in pseudo-rapidity between the two leptons (\deletall) measured by CMS~\cite{CMS:2018adi}. 

The SM predictions for conventional $t\bar t$ production and decay, referred to as the null hypothesis \hnull\ in our analysis, are obtained from fixed-order NNLO calculations as provided in~\cite{Czakon:2020qbd}, using the NNLO set of NNPDF3.1 parton distribution functions~\cite{Bertone:2017bme}. We then define our alternative hypothesis \halt\ as the coherent sum of the conventional and toponium contributions. The impact of the latter is simulated following the prescription of~\cite{Fuks:2024yjj}, implemented in \mg~\mgversion~\cite{Alwall:2014hca}, in which the leading-order matrix element for the $g g\to b\bar{\ell}\nu_\ell \bar{b}\ell'\bar{\nu}_{\ell'}$ process is re-weighted by the Green's function of the non-relativistic QCD Hamiltonian computed from a tree-level Coulomb potential. Parton-level events are then passed to \pythia~\pythiaversion~\cite{Bierlich:2022pfr} for parton showering and hadronisation. Next, event selection and observable definitions follow those of the ATLAS and CMS measurements as implemented within the \rivet~\rivetversion\ framework~\cite{Bierlich:2024vqo}, using the corresponding public routines identified by the \rivet codes \rivetroutinecodeATLAS\ and \rivetroutinecodeCMS.

\begin{figure}
	\centering \includegraphics[width=0.48\textwidth, angle=0]{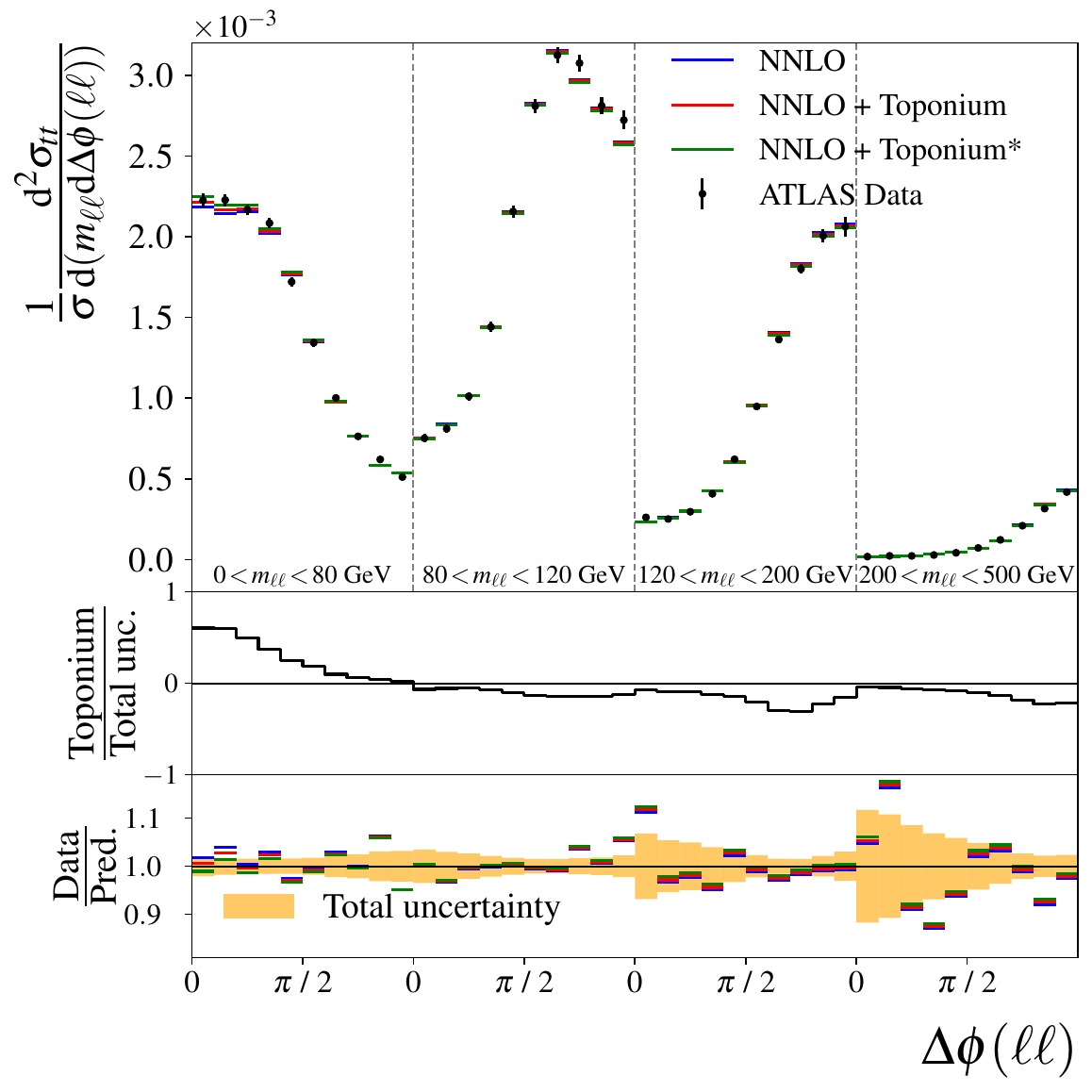}
	\caption{Data in the two-dimensional differential cross section of $\mll$ and $\delphill$~\cite{ATLAS:2019hau} are compared with predictions excluding (blue) and including a toponium component, using either the predicted (red) or best-fit (green) cross section. The upper panel shows the data and model predictions; the middle panel shows the ratio of the expected toponium signal to the experimental uncertainty in each bin; and the lower panel shows the ratio of the data to each prediction with the shaded band representing the total uncertainty on the data. \label{fig:atlas_mlldelphill}}
\end{figure}

In Figure~\ref{fig:atlas_mlldelphill} the two-dimensional differential cross section from~\cite{ATLAS:2019hau} is compared with model predictions that exclude (blue) or include a toponium component with a cross section obtained from the Green's-function re-weighting prescription of~\cite{Fuks:2024yjj} (red). The upper panel of the figure shows the data and the model curves. To illustrate the sensitivity of each bin to toponium effects, the middle panel displays the ratio of the toponium yield to the total uncertainty on the data. Finally, the lower panel of the figure displays the ratio of the data to each prediction, with the shaded band corresponding to the total uncertainty on the data.

Where toponium effects are expected to be largest, the inclusion of the toponium component to the model improves agreement between predictions and data. While suggestive, this feature alone is insufficient for a conclusive statement because conventional predictions in the same bins are sensitive to variations of SM parameters and modelling choices. We therefore assess consistency across multiple observables that retain discriminating power for toponium formation but are less affected by these modelling ambiguities. Similar comparisons for all considered one-dimensional observables are shown in the appendix.

\medskip

\textit{\textbf{Methodology and Analysis}} -- We quantify the presence of a toponium component in the six particle-level differential cross sections using complementary statistical measures. We define a $\chi^{2}$ test statistic under the assumption of Gaussian-distributed data,
\begin{equation}
  \chi^{2} = \mathrm{d}X^{T}\, V^{-1}\, \mathrm{d}X\,,
\end{equation}
where $\mathrm{d}X$ denotes the vector of absolute differences between the data and the prediction and $V$ is the data covariance matrix.
The covariance matrices for the CMS results are taken from the \textsc{HepData} database~\cite{hepdata.89307.v1}; for the ATLAS case, we instead adopt diagonal covariances matrices based on the bin uncertainties distributions. Although correlation matrices are provided in \cite{hepdata.91242}, we refrain from converting these into covariance matrices to avoid potential errors associated with subtleties in the definition of the correlation matrices. We subsequently encourage direct publication of covariance matrices for differential cross section data.
A likelihood-ratio test statistic is constructed as $\lambda = e^{-\frac{1}{2}\Delta \chi^{2}}$, with $\Delta \chi^{2}$ being the difference between the $\chi^{2}$ values computed for the conventional SM hypothesis \hnull\ and the alternative hypothesis \halt\ including toponium contributions. The distributions of $\lambda$ under \hnull\ and \halt\ are then obtained via pseudo-experiments assuming Gaussian data and using the covariance matrix to model inter-bin correlations. The code used for the  analysis is publicly available.\footnote{The codebase is released under a GNU General Public License at \gitlink.}

\begin{table}\renewcommand{\arraystretch}{1.5}\setlength{\tabcolsep}{5pt}
  \begin{tabular}{cc|cc|cc|c}
    \multirow{2}{*}{Observable} & \multirow{2}{*}{Exp.} & \multicolumn{2}{c|}{\hnull} & \multicolumn{2}{c|}{\halt} &\multirow{2}{*}{$B_{10}$} \\
     &  & $p$ & $Z$ & $p$ & $Z$ &  \\
    \hline
    $\frac{\mathrm{d}^{2} \sigma}{\mathrm{d}\mll\, \mathrm{d}\delphill}$ & ATLAS & $8\!\cdot\! 10^{-4}$ & 3 & 0.04 & 2 & 27 \\
    $\frac{\mathrm{d}\sigma}{\mathrm{d}\mll}$ & ATLAS & $4\!\cdot\! 10^{-5}$ & 4 & $9\!\cdot\! 10^{-4}$ & 3 & 20 \\
    $\frac{\mathrm{d}\sigma}{\mathrm{d}\mll}$ & CMS & 0.002 & 3 & 0.02 & 2 & 7 \\
    $\frac{\mathrm{d}\sigma}{\mathrm{d}\delphill}$ & ATLAS & 0.02 & 2 & 0.1 & 1 & 4 \\
    $\frac{d\sigma}{\mathrm{d}\delphill}$ & CMS & 0.3 & 0.5 & 0.3 & 0.4 & 1 \\
    $\frac{\mathrm{d}\sigma}{\mathrm{d}\deletall}$ & CMS & $5\!\cdot\! 10^{-4}$ & 3 & $0.003$ & 3 & 4 \\
  \end{tabular}
  \caption{Observed $p$-values, corresponding significances ($Z$) and Bayes factors ($B_{10}$) for the six measured differential cross sections under the SM-only (\hnull) and SM+toponium (\halt) hypotheses.  \label{tab:results}}
\end{table}

Table~\ref{tab:results} summarises the results for each observable, including the $p$-values, significances ($Z$) and Bayes factors $B_{10}$. Following the approach introduced in~\cite{Kass:1995loi, Fowlie:2024dgj, Fuks:2024qdt, Agin:2025vgn}, the Bayes factor reduces to the ratio of likelihood values in the absence of free parameters and provides a compact summary of the data’s preference for one hypothesis over the other: values of $B_{10} > 1$ indicate support for the alternative hypothesis, while values exceeding 10 correspond to strong evidence. For five of the six observables, the $p$-values under $\hnull$ fall below 0.05 ($Z \gtrsim 2$), indicating rejection of the conventional SM hypothesis in favour of the toponium-enhanced one at the 95\% confidence level. Nevertheless, the $p$-values ($Z$ significances) under \halt\ remain small (large), implying that the data are not yet fully described even after including toponium formation effects. However, the most sensitive observables, $\mathrm{d}^{2}\sigma / \mathrm{d}\mll\,\mathrm{d}\delphill$ and $\mathrm{d}\sigma / \mathrm{d}\mll$, yield Bayes factors exceeding 10. The only exception in supporting one hypothesis over the other, the CMS $\mathrm{d}\sigma / \mathrm{d}\delphill$ measurement, exhibits weak sensitivity owing to the limited impact of toponium formation on its shape within the restricted fiducial phase space of the measurement.

\begin{figure}
	\centering \includegraphics[width=0.45\textwidth]{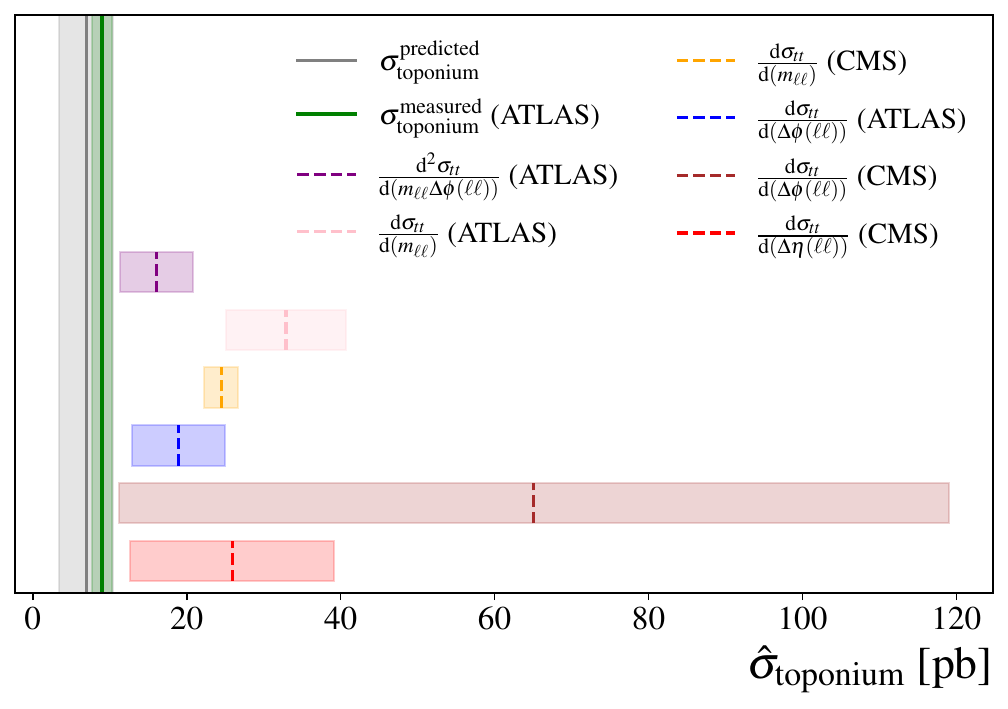}
	\caption{Values of the estimated toponium cross section $\hat{\sigma}_{\rm toponium}$ for each observable (dashed lines with shaded bands). The theoretical prediction from~\cite{Fuks:2024yjj} with a 50\% uncertainty is shown as a grey band, and the ATLAS detector-level measurement~\cite{ATLAS:2025kvb} is shown with a green band.
    \label{fig:summary}}
\end{figure}

We extend the analysis by determining the inclusive toponium cross section through a minimisation of the $\Delta\chi^{2}$ between the data and a model in which the toponium contribution is scaled by a free parameter before being added to the conventional prediction. The resulting values, $\hat{\sigma}_{\rm toponium}$, are shown in Figure~\ref{fig:summary} for each observable together with their 68\% confidence intervals, and can be compared with the theoretical prediction obtained from the Green's-function re-weighting prescription of~\cite{Fuks:2024yjj} including a 50\% uncertainty (grey band), and the measured detector-level cross section from ATLAS~\cite{ATLAS:2025kvb} (green band). The corresponding impact on each observable is additionally displayed by the green curves in Figure~\ref{fig:atlas_mlldelphill} and in the figures of the appendix. 

The extracted $\hat{\sigma}_{\rm toponium}$ values are broadly consistent across observables within one standard deviation, though they tend to exceed both the theoretical prediction and the ATLAS measurement. This pattern may reflect missing higher-order corrections in the conventional predictions, NNLO effects in top-quark decay not included in the LHC analyses or subtle differences in fiducial phase-space definitions. The potential for new physics explanations, as explored for instance in~\cite{Banik:2023vxa, Coloretti:2023yyq, Maltoni:2024wyh, Lu:2024twj, Llanes-Estrada:2024phk, Djouadi:2024lyv}, should therefore only be assessed after accounting for both these effects and toponium contributions, following the path described in~\cite{Flacke:2025dwk}.
Future particle-level measurements optimised for toponium sensitivity, released with public \textsc{Rivet} routines and associated covariance matrices as well as higher-order QCD and electroweak predictions, will allow these mild tensions to be probed with greater precision. 

\medskip

\textit{\textbf{Summary and Conclusions}} -- We have analysed six particle-level differential cross sections of $t \bar{t}$ production measured by the ATLAS and CMS collaborations to probe for evidence of toponium formation. The data were compared to conventional predictions at NNLO QCD accuracy and to predictions incorporating toponium contributions. In all observables, the inclusion of toponium formation improves the agreement between data and theory. In addition, a set of one-parameter $\chi^{2}$ minimisations was performed to extract the inclusive toponium production cross section associated with each distribution. The resulting estimates are broadly consistent across observables, though they show mild tension with both the theoretical prediction and a recent detector-level measurement from ATLAS. Taken together, the improved modelling and the consistency of the extracted cross-section estimates suggest that non-relativistic threshold effects beyond conventional perturbative predictions play an important role in describing LHC data. However, a combined determination of the significance and inclusive cross section across all observables is not currently feasible owing to the absence of publicly available information on correlations between distributions and between ATLAS and CMS data. A joint treatment of the two datasets, or a public release of inter-distribution and inter-experiment covariance matrices, would enable a definitive combined result and provide deeper insight into the presence of toponium in data. 

Given the complexity of $t \bar{t}$ production near threshold, further dedicated studies are required to disentangle possible toponium effects from modelling uncertainties or potential beyond-the-Standard-Model contributions. The growing availability of double- and triple-differential $t\bar t$ measurements, along with corresponding predictions and public Monte Carlo implementations of toponium production, will enable more definitive tests. Within the experimental measurements, machine-learning techniques trained to discriminate toponium effects from modelling variations are expected to play an important role. 

%%%%%%%%%%%%%%%%%%%%%%%%%%%%%%%%%%
\section*{Acknowledgements}
We deeply thank Y.~Afik, K.~Hagiwara and B.~Ravina for valuable comments on the manuscript. The work of BF was supported in part by Grant ANR-21-CE31-0013 (Project DMwithLLPatLHC) from the French \emph{Agence Nationale de la Recherche} (ANR). The work of AH was supported in part by a bursary from the University of Cape Town.

%%%%%%%%%%%%%%%%%%%%%%%%%%%%%%%%%%

\appendix \section{Appendix: Additional Figures}

\begin{figure*}
  \centering
  \includegraphics[width=0.48\textwidth]{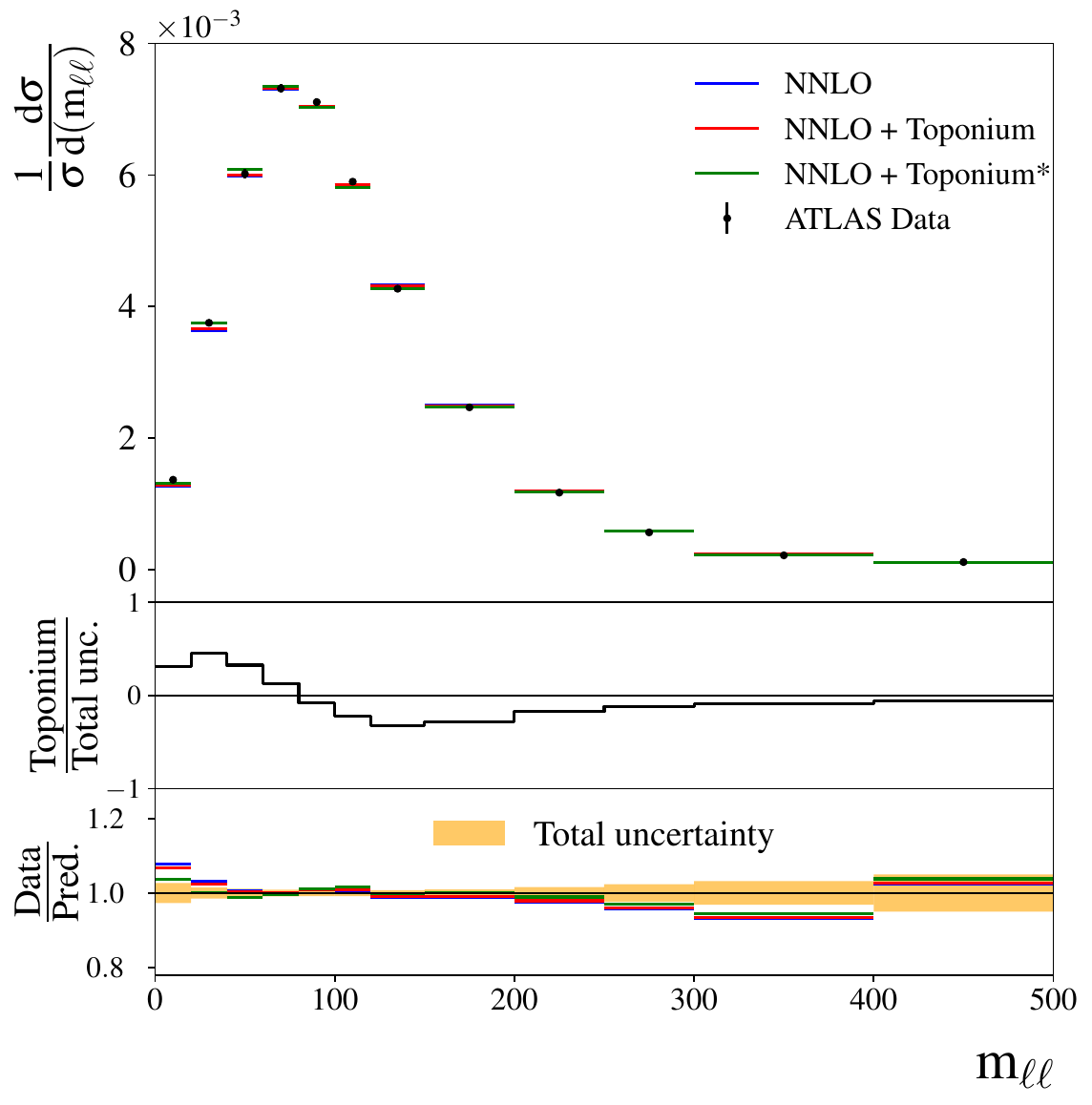} 
  \hfill
  \includegraphics[width=0.48\textwidth]{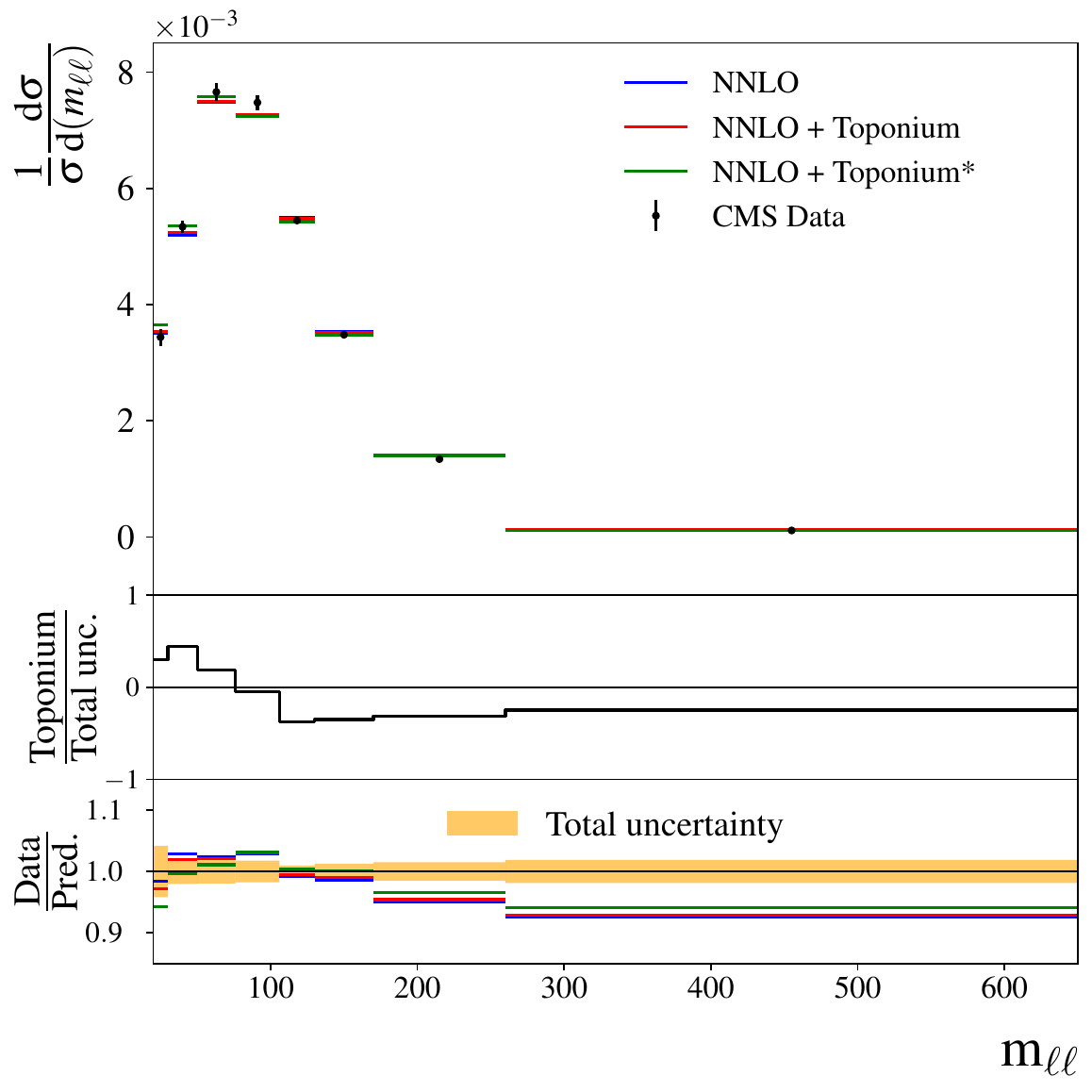}
  \caption{Same as Figure~\ref{fig:atlas_mlldelphill} but for $\frac{\mathrm{d}\sigma}{\mathrm{d}\mll}$ as measured from \cite{ATLAS:2019hau} (left) and \cite{CMS:2018adi} (right). \label{fig:mll}}
\end{figure*}

\begin{figure}
  \centering
  \includegraphics[width=0.48\textwidth]{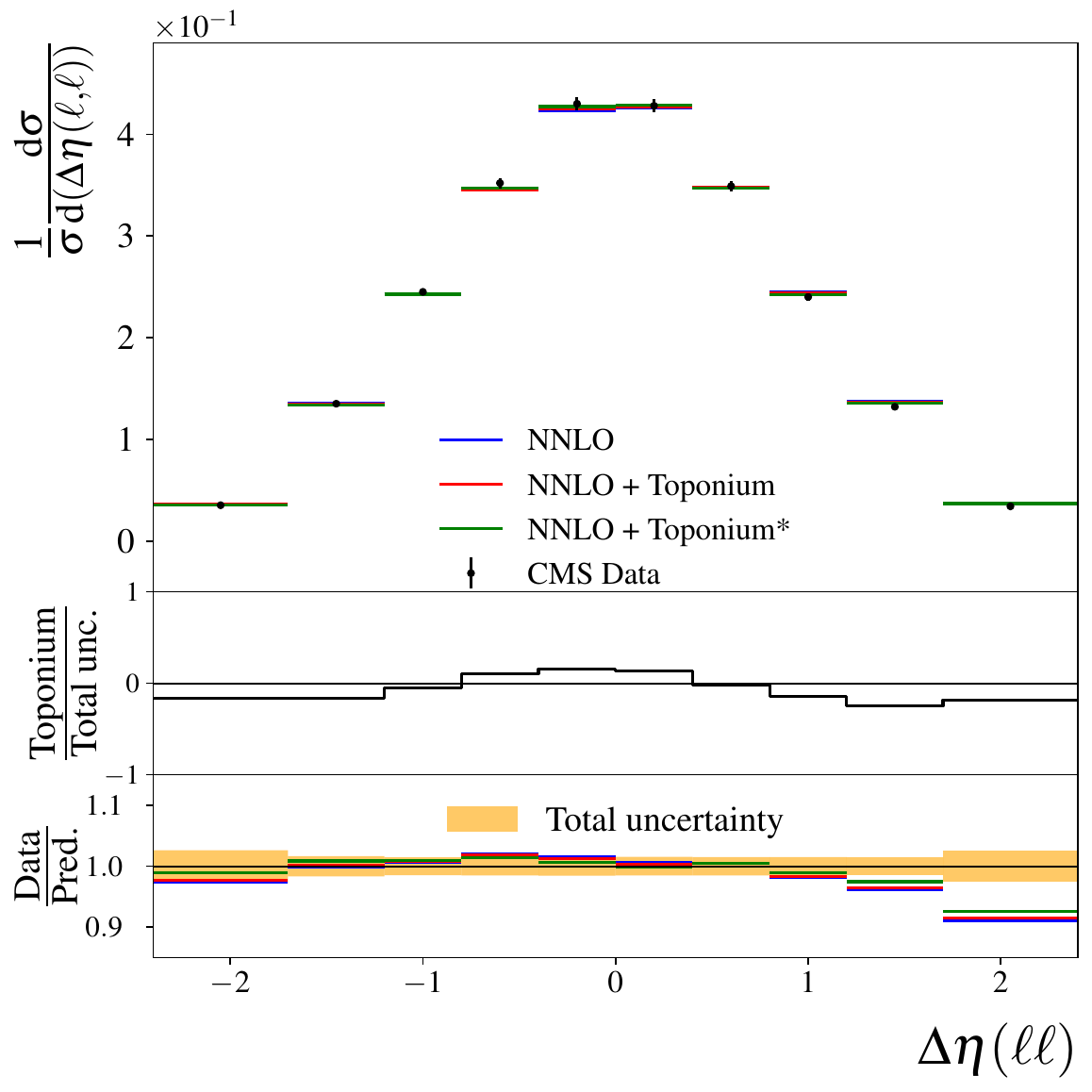}
  \caption{Same as Figure~\ref{fig:atlas_mlldelphill} but for $\frac{\mathrm{d}\sigma}{\mathrm{d}\deletall}$ as measured from~\cite{CMS:2018adi}. \label{fig:deletacms}}
\end{figure}

Figures~\ref{fig:mll}-\ref{fig:deletacms} show comparisons between the measured data, the conventional predictions and those including toponium contributions for the one-dimensional differential cross sections analysed. In each figure, the upper panel displays the measured data together with the theoretical predictions corresponding to the three signal treatments: excluding toponium contributions (blue), including them with the predicted cross section (red) or with the best-fit value (green). The middle panel shows the ratio of the expected toponium signal in each bin to the total uncertainty, while the lower panel shows the ratio of the measured data to each prediction with the shaded band indicating the total uncertainty.

\begin{figure*}
  \centering
  \includegraphics[width=0.48\textwidth]{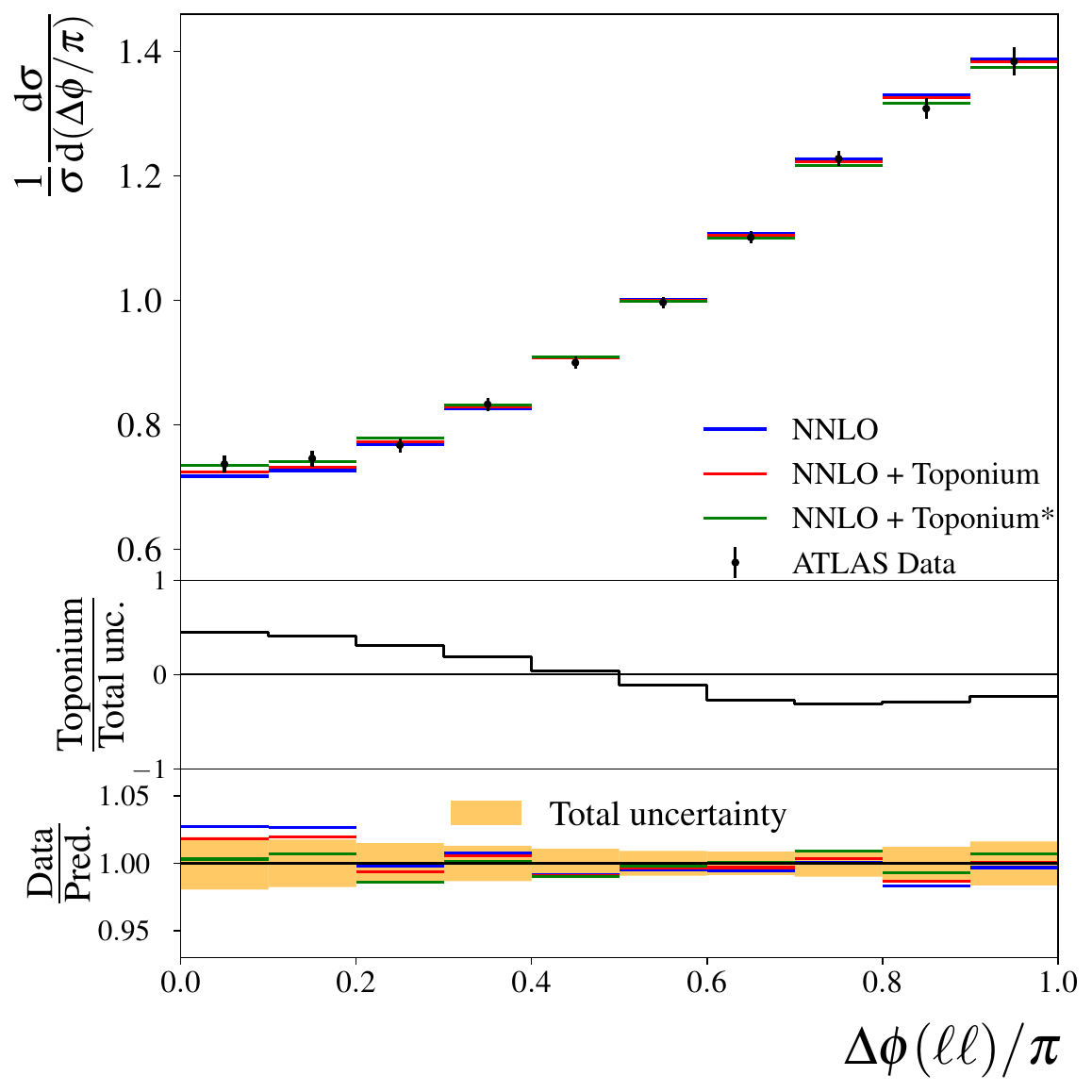}
  \hfill
  \includegraphics[width=0.48\textwidth]{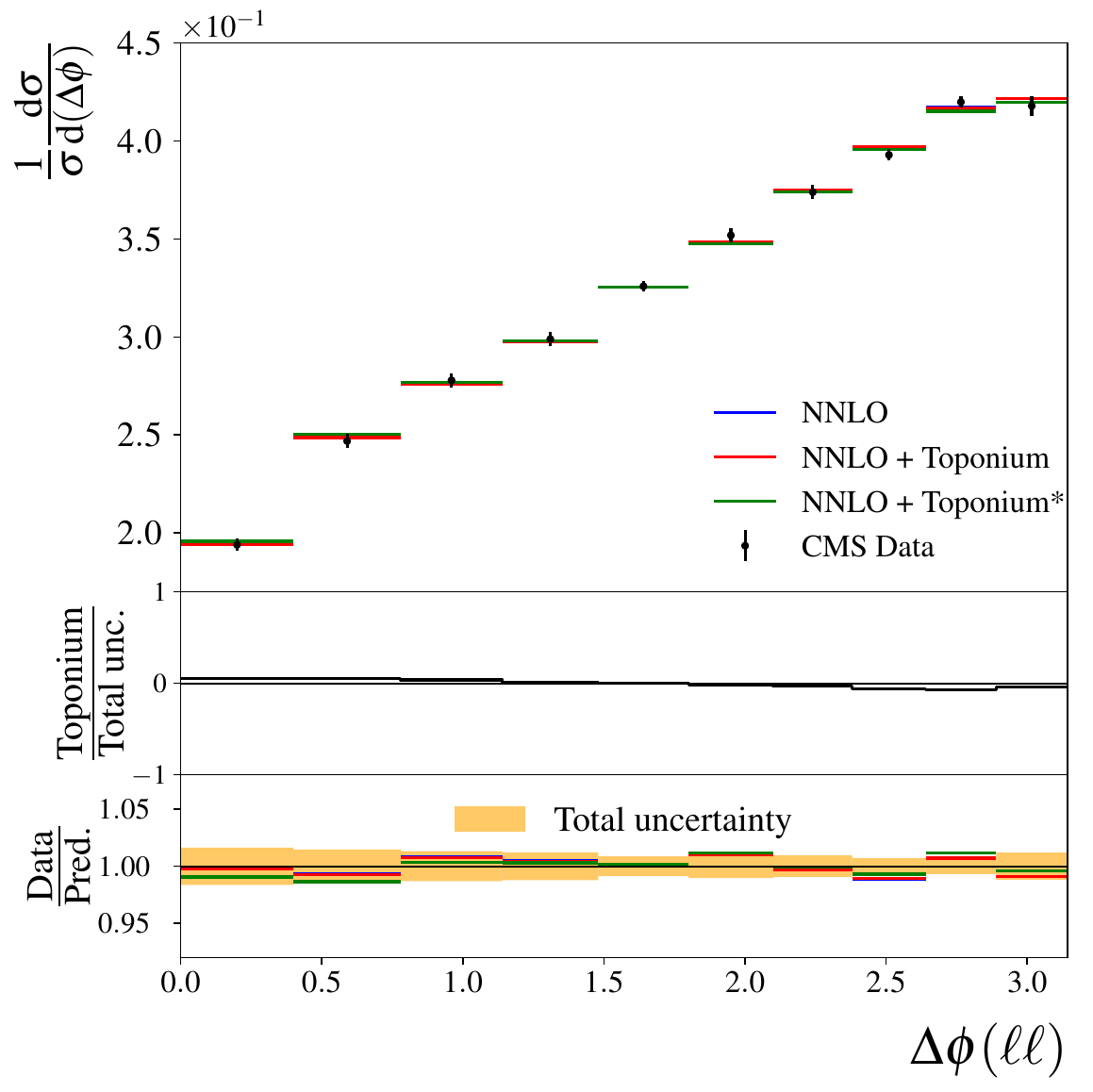}
  \caption{Same as Figure~\ref{fig:atlas_mlldelphill} but for $\frac{\mathrm{d}\sigma}{\mathrm{d}\delphill}$ as measured from \cite{ATLAS:2019hau} (left) and \cite{CMS:2018adi} (right).}
\end{figure*}

\bibliographystyle{JHEP}
\bibliography{toponium}

\providecommand{\href}[2]{#2}\begingroup\raggedright\begin{thebibliography}{10}

\bibitem{Fadin:1987wz}
V.S.~Fadin and V.A.~Khoze, \emph{{Threshold Behavior of Heavy Top Production in
  e+ e- Collisions}}, {\emph{JETP Lett.} {\bfseries 46} (1987) 525}.

\bibitem{Fadin:1990wx}
V.S.~Fadin, V.A.~Khoze and T.~Sjostrand, \emph{{On the Threshold Behavior of
  Heavy Top Production}}, \href{https://doi.org/10.1007/BF01614696}{\emph{Z.
  Phys. C} {\bfseries 48} (1990) 613}.

\bibitem{Hagiwara:2008df}
K.~Hagiwara, Y.~Sumino and H.~Yokoya, \emph{{Bound-state Effects on Top Quark
  Production at Hadron Colliders}},
  \href{https://doi.org/10.1016/j.physletb.2008.07.006}{\emph{Phys. Lett. B}
  {\bfseries 666} (2008) 71} [\href{https://arxiv.org/abs/0804.1014}{{\ttfamily
  0804.1014}}].

\bibitem{Sumino:2010bv}
Y.~Sumino and H.~Yokoya, \emph{{Bound-state effects on kinematical
  distributions of top quarks at hadron colliders}},
  \href{https://doi.org/10.1007/JHEP09(2010)034}{\emph{JHEP} {\bfseries 09}
  (2010) 034} [\href{https://arxiv.org/abs/1007.0075}{{\ttfamily 1007.0075}}].

\bibitem{Fuks:2021xje}
B.~Fuks, K.~Hagiwara, K.~Ma and Y.-J.~Zheng, \emph{{Signatures of toponium
  formation in LHC run 2 data}},
  \href{https://doi.org/10.1103/PhysRevD.104.034023}{\emph{Phys. Rev. D}
  {\bfseries 104} (2021) 034023}
  [\href{https://arxiv.org/abs/2102.11281}{{\ttfamily 2102.11281}}].

\bibitem{Fuks:2024yjj}
B.~Fuks, K.~Hagiwara, K.~Ma and Y.-J.~Zheng, \emph{{Simulating toponium
  formation signals at the LHC}},
  \href{https://doi.org/10.1140/epjc/s10052-025-13853-3}{\emph{Eur. Phys. J. C}
  {\bfseries 85} (2025) 157}
  [\href{https://arxiv.org/abs/2411.18962}{{\ttfamily 2411.18962}}].

\bibitem{Ju:2020otc}
W.-L.~Ju, G.~Wang, X.~Wang, X.~Xu, Y.~Xu and L.L.~Yang, \emph{{Top quark pair
  production near threshold: single/double distributions and mass
  determination}}, \href{https://doi.org/10.1007/JHEP06(2020)158}{\emph{JHEP}
  {\bfseries 06} (2020) 158}
  [\href{https://arxiv.org/abs/2004.03088}{{\ttfamily 2004.03088}}].

\bibitem{Garzelli:2024uhe}
M.V.~Garzelli, G.~Limatola, S.O.~Moch, M.~Steinhauser and O.~Zenaiev,
  \emph{{Updated predictions for toponium production at the LHC}},
  \href{https://doi.org/10.1016/j.physletb.2025.139532}{\emph{Phys. Lett. B}
  {\bfseries 866} (2025) 139532}
  [\href{https://arxiv.org/abs/2412.16685}{{\ttfamily 2412.16685}}].

\bibitem{Nason:2025hix}
P.~Nason, E.~Re and L.~Rottoli, \emph{{Spin Correlations in $t{\bar t}$
  Production and Decay at the LHC in QCD Perturbation Theory}},
  \href{https://arxiv.org/abs/2505.00096}{{\ttfamily 2505.00096}}.

\bibitem{Fuks:2025wtq}
B.~Fuks, K.~Hagiwara, K.~Ma, L.~Munoz-Aillaud and Y.-J.~Zheng, \emph{{Prospects
  for toponium formation at the LHC in the single-lepton mode}},
  \href{https://arxiv.org/abs/2509.03596}{{\ttfamily 2509.03596}}.

\bibitem{ATLAS:2019hau}
{\scshape ATLAS} collaboration, \emph{{Measurement of the $t\bar{t}$ production
  cross-section and lepton differential distributions in $e\mu $ dilepton
  events from $pp$ collisions at $\sqrt{s}=13\,\text {TeV}$ with the ATLAS
  detector}}, \href{https://doi.org/10.1140/epjc/s10052-020-7907-9}{\emph{Eur.
  Phys. J. C} {\bfseries 80} (2020) 528}
  [\href{https://arxiv.org/abs/1910.08819}{{\ttfamily 1910.08819}}].

\bibitem{ATLAS:2019zrq}
{\scshape ATLAS} collaboration, \emph{{Measurements of top-quark pair spin
  correlations in the $e\mu$ channel at $\sqrt{s} = 13$ TeV using $pp$
  collisions in the ATLAS detector}},
  \href{https://doi.org/10.1140/epjc/s10052-020-8181-6}{\emph{Eur. Phys. J. C}
  {\bfseries 80} (2020) 754}
  [\href{https://arxiv.org/abs/1903.07570}{{\ttfamily 1903.07570}}].

\bibitem{CMS:2018adi}
{\scshape CMS} collaboration, \emph{{Measurements of $\mathrm{t\overline{t}}$
  differential cross sections in proton-proton collisions at $\sqrt{s}=$ 13 TeV
  using events containing two leptons}},
  \href{https://doi.org/10.1007/JHEP02(2019)149}{\emph{JHEP} {\bfseries 02}
  (2019) 149} [\href{https://arxiv.org/abs/1811.06625}{{\ttfamily
  1811.06625}}].

\bibitem{CMS:2024ybg}
{\scshape CMS} collaboration, \emph{{Differential cross section measurements
  for the production of top quark pairs and of additional jets using dilepton
  events from pp collisions at $\sqrt{s}$ = 13 TeV}},
  \href{https://doi.org/10.1007/JHEP02(2025)064}{\emph{JHEP} {\bfseries 02}
  (2025) 064} [\href{https://arxiv.org/abs/2402.08486}{{\ttfamily
  2402.08486}}].

\bibitem{ATLAS:2023fsd}
{\scshape ATLAS} collaboration, \emph{{Observation of quantum entanglement with
  top quarks at the ATLAS detector}},
  \href{https://doi.org/10.1038/s41586-024-07824-z}{\emph{Nature} {\bfseries
  633} (2024) 542} [\href{https://arxiv.org/abs/2311.07288}{{\ttfamily
  2311.07288}}].

\bibitem{CMS:2024pts}
{\scshape CMS} collaboration, \emph{{Observation of quantum entanglement in top
  quark pair production in proton{\textendash}proton collisions at $\sqrt{s} =
  13$ TeV}}, \href{https://doi.org/10.1088/1361-6633/ad7e4d}{\emph{Rept. Prog.
  Phys.} {\bfseries 87} (2024) 117801}
  [\href{https://arxiv.org/abs/2406.03976}{{\ttfamily 2406.03976}}].

\bibitem{CMS:2024zkc}
{\scshape CMS} collaboration, \emph{{Measurements of polarization and spin
  correlation and observation of entanglement in top quark pairs using
  lepton+jets events from proton-proton collisions at s=13{\,}{\,}TeV}},
  \href{https://doi.org/10.1103/PhysRevD.110.112016}{\emph{Phys. Rev. D}
  {\bfseries 110} (2024) 112016}
  [\href{https://arxiv.org/abs/2409.11067}{{\ttfamily 2409.11067}}].

\bibitem{CMS:2025kzt}
{\scshape CMS} collaboration, \emph{{Observation of a pseudoscalar excess at
  the top quark pair production threshold}},
  \href{https://doi.org/10.1088/1361-6633/adf7d3}{\emph{Rept. Prog. Phys.}
  {\bfseries 88} (2025) 087801}
  [\href{https://arxiv.org/abs/2503.22382}{{\ttfamily 2503.22382}}].

\bibitem{CMS:2025dzq}
{\scshape CMS} collaboration, \emph{{Search for heavy pseudoscalar and scalar
  bosons decaying to a top quark pair in proton-proton collisions at $\sqrt{s}$
  = 13 TeV}},  \href{https://arxiv.org/abs/2507.05119}{{\ttfamily 2507.05119}}.

\bibitem{ATLAS:2025kvb}
{\scshape ATLAS} collaboration, \emph{{Observation of a cross-section
  enhancement near the $t\bar{t}$ production threshold in $\sqrt{s}=13$ TeV
  $pp$ collisions with the ATLAS detector}},  ATLAS-CONF-2025-008.

\bibitem{Czakon:2020qbd}
M.~Czakon, A.~Mitov and R.~Poncelet, \emph{{NNLO QCD corrections to leptonic
  observables in top-quark pair production and decay}},
  \href{https://doi.org/10.1007/JHEP05(2021)212}{\emph{JHEP} {\bfseries 05}
  (2021) 212} [\href{https://arxiv.org/abs/2008.11133}{{\ttfamily
  2008.11133}}].

\bibitem{Bertone:2017bme}
{\scshape NNPDF} collaboration, \emph{{Illuminating the photon content of the
  proton within a global PDF analysis}},
  \href{https://doi.org/10.21468/SciPostPhys.5.1.008}{\emph{SciPost Phys.}
  {\bfseries 5} (2018) 008} [\href{https://arxiv.org/abs/1712.07053}{{\ttfamily
  1712.07053}}].

\bibitem{Alwall:2014hca}
J.~Alwall, R.~Frederix, S.~Frixione, V.~Hirschi, F.~Maltoni, O.~Mattelaer
  et~al., \emph{{The automated computation of tree-level and next-to-leading
  order differential cross sections, and their matching to parton shower
  simulations}}, \href{https://doi.org/10.1007/JHEP07(2014)079}{\emph{JHEP}
  {\bfseries 07} (2014) 079} [\href{https://arxiv.org/abs/1405.0301}{{\ttfamily
  1405.0301}}].

\bibitem{Bierlich:2022pfr}
C.~Bierlich et~al., \emph{{A comprehensive guide to the physics and usage of
  PYTHIA 8.3}},
  \href{https://doi.org/10.21468/SciPostPhysCodeb.8}{\emph{SciPost Phys.
  Codeb.} {\bfseries 2022} (2022) 8}
  [\href{https://arxiv.org/abs/2203.11601}{{\ttfamily 2203.11601}}].

\bibitem{Bierlich:2024vqo}
C.~Bierlich, A.~Buckley, J.M.~Butterworth, C.~Gutschow, L.~Lonnblad, T.~Procter
  et~al., \emph{{Robust independent validation of experiment and theory: Rivet
  version 4 release note}},
  \href{https://doi.org/10.21468/SciPostPhysCodeb.36}{\emph{SciPost Phys.
  Codeb.} {\bfseries 36} (2024) 1}
  [\href{https://arxiv.org/abs/2404.15984}{{\ttfamily 2404.15984}}].

\bibitem{hepdata.89307.v1}
{CMS Collaboration}, \emph{{Measurements of $\mathrm{t\overline{t}}$
  differential cross sections in proton-proton collisions at $\sqrt{s} =$ 13
  TeV using events containing two leptons (Version 1)}},
  \url{https://doi.org/10.17182/hepdata.89307.v1}.

\bibitem{hepdata.91242}
{ATLAS Collaboration}, \emph{{Measurement of the $t\bar{t}$ production
  cross-section and lepton differential distributions in $e\mu $ dilepton
  events from $pp$ collisions at $\sqrt{s}=13\,\text {TeV}$ with the ATLAS
  detector (Version 1)}},  \url{https://doi.org/10.17182/hepdata.91242.v1}.

\bibitem{Kass:1995loi}
R.E.~Kass and A.E.~Raftery, \emph{{Bayes Factors}},
  \href{https://doi.org/10.1080/01621459.1995.10476572}{\emph{J. Am. Statist.
  Assoc.} {\bfseries 90} (1995) 773}.

\bibitem{Fowlie:2024dgj}
A.~Fowlie, \emph{{The Bayes factor surface for searches for new physics}},
  \href{https://doi.org/10.1140/epjc/s10052-024-12792-9}{\emph{Eur. Phys. J. C}
  {\bfseries 84} (2024) 426}
  [\href{https://arxiv.org/abs/2401.11710}{{\ttfamily 2401.11710}}].

\bibitem{Fuks:2024qdt}
B.~Fuks, M.D.~Goodsell and T.~Murphy, \emph{{Monojets from compressed weak
  frustrated dark matter}},
  \href{https://doi.org/10.1103/PhysRevD.111.055010}{\emph{Phys. Rev. D}
  {\bfseries 111} (2025) 055010}
  [\href{https://arxiv.org/abs/2409.03014}{{\ttfamily 2409.03014}}].

\bibitem{Agin:2025vgn}
D.~Agin, B.~Fuks, M.D.~Goodsell and T.~Murphy, \emph{{A joint explanation for
  the soft lepton and monojet LHC excesses in the wino-bino model}},
  \href{https://doi.org/10.1140/epjc/s10052-025-14886-4}{\emph{Eur. Phys. J. C}
  {\bfseries 85} (2025) 1145}
  [\href{https://arxiv.org/abs/2506.21676}{{\ttfamily 2506.21676}}].

\bibitem{Banik:2023vxa}
S.~Banik, G.~Coloretti, A.~Crivellin and B.~Mellado, \emph{{Uncovering new
  Higgses in the LHC analyses of differential $ t\overline{t} $ cross
  sections}}, \href{https://doi.org/10.1007/JHEP01(2025)155}{\emph{JHEP}
  {\bfseries 01} (2025) 155}
  [\href{https://arxiv.org/abs/2308.07953}{{\ttfamily 2308.07953}}].

\bibitem{Coloretti:2023yyq}
G.~Coloretti, A.~Crivellin and B.~Mellado, \emph{{Combined explanation of LHC
  multilepton, diphoton, and top-quark excesses}},
  \href{https://doi.org/10.1103/PhysRevD.110.073001}{\emph{Phys. Rev. D}
  {\bfseries 110} (2024) 073001}
  [\href{https://arxiv.org/abs/2312.17314}{{\ttfamily 2312.17314}}].

\bibitem{Maltoni:2024wyh}
F.~Maltoni, D.~Pagani and S.~Tentori, \emph{{Top-quark pair production as a
  probe of light top-philic scalars and anomalous Higgs interactions}},
  \href{https://doi.org/10.1007/JHEP09(2024)098}{\emph{JHEP} {\bfseries 09}
  (2024) 098} [\href{https://arxiv.org/abs/2406.06694}{{\ttfamily
  2406.06694}}].

\bibitem{Lu:2024twj}
C.-T.~Lu, K.~Cheung, D.~Kim, S.~Lee and J.~Song, \emph{{Can a pseudoscalar with
  a mass of 365 GeV in two-Higgs-doublet models explain the CMS
  tt{\textasciimacron} excess?}},
  \href{https://doi.org/10.1016/j.physletb.2024.139121}{\emph{Phys. Lett. B}
  {\bfseries 859} (2024) 139121}
  [\href{https://arxiv.org/abs/2410.08609}{{\ttfamily 2410.08609}}].

\bibitem{Llanes-Estrada:2024phk}
F.J.~Llanes-Estrada, \emph{{Ensuring that toponium is glued, not nailed}},
  \href{https://doi.org/10.1016/j.physletb.2025.139510}{\emph{Phys. Lett. B}
  {\bfseries 866} (2025) 139510}
  [\href{https://arxiv.org/abs/2411.19180}{{\ttfamily 2411.19180}}].

\bibitem{Djouadi:2024lyv}
A.~Djouadi, J.~Ellis and J.~Quevillon, \emph{{Contrasting pseudoscalar Higgs
  and toponium states at the LHC and beyond}},
  \href{https://doi.org/10.1016/j.physletb.2025.139583}{\emph{Phys. Lett. B}
  {\bfseries 866} (2025) 139583}
  [\href{https://arxiv.org/abs/2412.15138}{{\ttfamily 2412.15138}}].

\bibitem{Flacke:2025dwk}
T.~Flacke, B.~Fuks, D.~Kim, J.~Kim, S.J.~Lee and L.~Munoz-Aillaud, \emph{{New
  physics in toponium's shadow?}},
  \href{https://arxiv.org/abs/2512.03220}{{\ttfamily 2512.03220}}.

\end{thebibliography}\endgroup

\end{document}